\documentclass[10pt]{article}

\usepackage{amsmath}
\usepackage{amssymb}
\usepackage[section]{placeins}
\usepackage{graphicx}
\usepackage{cite}
\usepackage{color} 
\usepackage{epstopdf}
\usepackage{tikz}
\usepackage{mathtools}
\usepackage{microtype}
\DisableLigatures[f]{encoding = *, family = * }

\usepackage[labelfont=bf,labelsep=period,justification=raggedright]{caption}

\renewcommand{\vec}[1]{\mathbf{#1}}

\makeatletter
\renewcommand{\@biblabel}[1]{\quad#1.}
\makeatother

\date{}

\newcommand{\betti}{\mathsf{b}}
\newcommand{\proximity}{\varepsilon}

\renewcommand{\exp}[1]{\mathrm{e}^{#1}}

\newcommand{\C}{\mathsf{C}}

\newcommand{\B}{\mathsf{B}}
\renewcommand{\H}{\mathsf{H}}
\newcommand{\Z}{\mathsf{Z}}
\newcommand{\ZZ}{\mathbb{Z}}

\newcommand{\im}{\mathsf{im}}
\renewcommand{\ker}{\mathsf{ker}}
\newcommand{\Order}{\mathcal{O}}

\begin{document}

\begin{flushleft}
{\Large
\textbf{Topological Data Analysis of Biological Aggregation Models}
}
\\
Chad M. Topaz$^{1,\ast}$,
Lori Ziegelmeier$^{1}$, 
Tom Halverson$^{1}$, 
\\
\bf{1} Department of Mathematics, Statistics, and Computer Science, Macalester College, Saint Paul, Minnesota, United States of America
\\
$\ast$ E-mail: ctopaz@macalester.edu
\end{flushleft}

\section*{Abstract}

We apply tools from topological data analysis to two mathematical models inspired by biological aggregations such as bird flocks, fish schools, and insect swarms. Our data consists of numerical simulation output from the models of Vicsek and D'Orsogna. These models are dynamical systems describing the movement of agents who interact via alignment, attraction, and/or repulsion. Each simulation time frame is a point cloud in position-velocity space. We analyze the topological structure of these point clouds, interpreting the persistent homology by calculating the first few Betti numbers. These Betti numbers count connected components, topological circles, and trapped volumes present in the data. To interpret our results, we introduce a visualization that displays Betti numbers over simulation time and topological persistence scale. We compare our topological results to order parameters typically used to quantify the global behavior of aggregations, such as polarization and angular momentum. The topological calculations reveal events and structure not captured by the order parameters.

\section*{Introduction}

Biological aggregations are groups of organisms such as fish schools, bird flocks, insect swarms, and mammal herds \cite{ParHam1997,OkuLev2001, CamDenFra2001}. Social interactions between members can play a crucial role in the formation and behavior of these groups \cite{ParEde1999, Sum2010, CouKraJam2002}. Social interactions are behaviors like attraction, repulsion, and alignment, which are activated when one organism senses another via sight, sound, smell, touch, or perhaps some combination of senses~\cite{EftVriLew2007}. Aggregations take on a vast array of morphologies: advancing fronts of running wildebeest, branched dendritic structures of bacteria, tornado-like vortices of swimming anchovy, and much more. Beyond serving as examples of emergent pattern formation, organisms moving in groups can affect resource consumption, disease transmission, and at the longest spatiotemporal scales, evolution itself \cite{OkuGruEde2001}. Beyond the realm of biology, the understanding of biological aggregations has inspired applications from computer algorithms to robotic self-assembly \cite{Pas2005}.

Quantitative understanding of aggregations has been developed in part through mathematical modeling. Modeling of aggregations dates back (at least) to the 1950s with the seminal work of \cite{Bre1954}, which describes the motion of individual fish as particles obeying Newton's law. The forces in the model are social forces between two fish -- namely attraction and repulsion -- and are described by a simple functional form dependent on the distance between individuals, akin to gravitational or intermolecular forces in physics. Since then, hundreds of aggregation models have been created; some of the most well-studied include \cite{VicCziBen1995, CouKraJam2002, ChuDOrMar2007}. The common approach is to envision organisms as point particles with motion laws that are first or second order in time, and with behavioral rules that are some combination of self-propulsion and/or social alignment, attraction, and repulsion. An alternative modeling approach, also the subject of a rich literature, is to treat a sufficiently large population as a continuum and describe its dynamics with a partial integrodifferential equation as in, \emph{e.g.}, \cite{MogEde1999, TopBerLew2006, BerTop2011, BerTop2013, PotMokLew2014}. The discrete and continuum models that we have mentioned here are largely phenomenological. They are minimal models inspired to greater or lesser degree by biological observation, and their minimalism allows one to understand the basic behaviors necessary or sufficient for a particular type of aggregation phenomenon.

Quantitative understanding of aggregations has also been developed through the exploration and modeling of rich data sets measured in the field or in experiment. This type of study has a much more recent history, as the technology necessary to gather and process large, accurate data sets did not exist several decades ago. Notable examples include data-based modeling of starlings \cite{BalCalCan2008}, ducks \cite{LukLiEde2010}, aphids \cite{NilPaiWar2013}, golden shiner fish \cite{TunKatIoa2013}, and desert locusts \cite{BazBarHal2012}. In data based studies, one typically collects time  series of organisms' positions, and possibly, velocities. One may use this data in two different ways: to infer the rules for motion that each individual follows and/or to characterize the collective dynamics of the group.

In a classical approach to characterizing collective dynamics, one begins with $N$ organisms' positions $\vec{x}_i$ and velocities $\vec{v}_i$, either from biological observation or numerical simulation of a model. One then calculates some global metric hoped to give insight into macroscopic dynamics. For instance, \cite{CouKraJam2002} simulates discrete swarmers who interact via attraction, repulsion, and alignment, and measures the group polarization $P$ and angular momentum $M_{ang}$,
\begin{equation}
\label{eq:metrics1}
P = \left| \frac{\sum_{i=1}^N \vec{v}_i}{{\sum_{i=1}^N |\vec{v}_i}|} \right|, \quad M_{ang} = \left| \frac{\sum_{i=1}^N \vec{r}_i \times \vec{v}_i}{\sum_{i=1}^N |\vec{r}_i| |\vec{v}_i|} \right|,
\end{equation}
where $\vec{r}_i = \vec{x}_i - \vec{x}_{cm}$  and $\vec{x}_{cm}$ is the center of mass of the group. By varying parameters that control the social interactions and plotting $P$ and $M_{ang}$, \cite{CouKraJam2002} identifies different regimes of behavior, including swarming, motion on a torus, and a highly parallel (polarized) group. A different aggregation model, from \cite{ChuDOrMar2007}, can produce a rotating annulus of individuals (a ``single mill'') or superposed, counterrotating annuli (a ``double mill''). Metrics $P$ and $M_{ang}$ cannot distinguish a single from a double mill, so \cite{ChuDOrMar2007} introduces the absolute angular momentum
\begin{equation}
\label{eq:metrics2}
M_{abs} = \left| \frac{\sum_{i=1}^N |\vec{r}_i \times \vec{v}_i|}{\sum_{i=1}^N |\vec{r}_i| |\vec{v}_i|} \right|,
\end{equation}
whose numerator differs from $M_{ang}$. Together, $M_{ang}$ and $M_{abs}$ can distinguish single from double mills. Other examples of metrics include the average number of neighbors with whom an individual interacts (which requires knowledge of interaction rules) and the mean distance to nearest neighbor \cite{HueAld2008}.

Our discussion here provides a sampling of metrics in the literature. Most are inspired by order parameters from physics, and many have been constructed \emph{a posteriori}, based on knowledge of the dynamic whose detection was desired. In our present work, we explore whether topology offers a natural way to characterize collective behavior. In brief, we use the methods of topological data analysis to compute the persistent homology of spatiotemporal aggregation data sets arising from numerical simulation of models. We introduce a new visualization to explore homological persistence over spatial scales and over time. As we will explain at length, this visualization displays a data point cloud's Betti numbers in a contour diagram. We refer to this plot as the Contour Realization Of Computed $k$-dimensional hole Evolution in the Rips complex (CROCKER). We show that our topological analysis reveals dynamical events not captured in time series of the classically studied order parameters. Our primary goal is to demonstrate the utility of topological data analysis for biological aggregations and similar applications.

We note that the term topology has been invoked in the aggregation literature in a different sense than we use it here. In both the biological modeling literature and the robotics literature, ``topology'' sometimes refers to the coupling scheme between agents, that is, by which members of the group a given individual is influenced \cite{OlfMur2004,BalCalCan2008}. For instance, in some models or algorithms, a given agent might be influenced by the closest agent, or the closest two agents, or the closest few agents within an agent's field of view. In this sense, topology is an input of the model. In a different sense, we are using topological tools to analyze the outputs of a model.

The rest of this paper is organized as follows. We begin with an overview of persistent homology. This discussion aims to present some of the key concepts to a mathematical reader unfamiliar with algebraic topology. Brief descriptions of computational methods and our data visualization follow. Then, we proceed to topological analyses of the models of Vicsek, \emph{et al.} \cite{VicCziBen1995} and D'Orsogna, \emph{et al.} \cite{DOrChuBer2006} before concluding.

\section*{Topological Data Analysis and Persistent Homology}
\label{sec:homology}

Homology is a tool from algebraic topology that measures the features of a topological space such as an annulus, sphere, torus, or more complicated surface or manifold.  In particular, homology can distinguish these spaces from one another by quantifying their connected components, topological circles, trapped volumes, and so forth. A finite set of data points can be viewed as a (noisy) sampling from an underlying topological space. One can measure the homology of the data by creating connections between proximate data points, varying the scale over which these connections are made, and looking for features that persist across scales. This is called persistent homology. Persistent homology has been used in a wide array of applications to uncover the topological structure of data, including neuroscience, language processing, natural images, signal analysis, bioinformatics, computer vision, and sensor networks \cite{SinMemIsh2008, Zhu2013, CarIshDe-2008, PerHar2013, KasZomPar2007,FreChe2009, SilGhr2007}.

We explain these ideas in greater detail for the remainder of this section. Our discussions in the first two subsections below recapitulate presentations in texts such as \cite{Hat2002,Cro2006}.

\subsection*{Forming a Simplicial Complex}

\begin{figure}[t]
\centering
\includegraphics[width=0.9\textwidth]{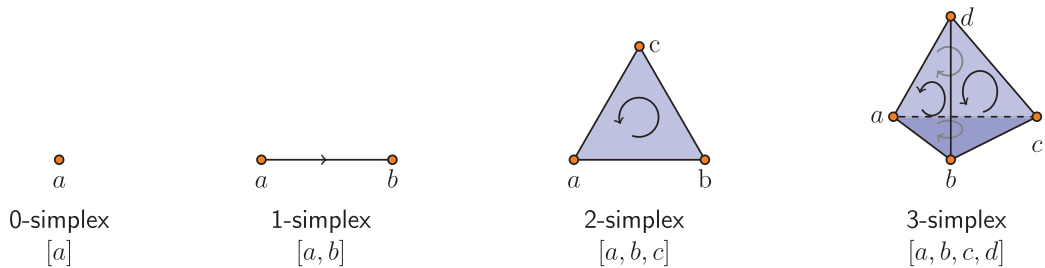} 
\caption{\label{fig:simplices}
{\bf Oriented $k$-simplices for $k = 0, 1, 2, 3$.} These $k$-simplices are the building blocks used to construct a simplicial complex from a point cloud of data.
}
\hrule
\end{figure}

To build a global object from a discrete set of $N$ data points, we construct  a \emph{simplicial complex} $S$.  A simplicial complex is a set consisting of a finite collection of $k$-simplices (simple pieces), where a 0-simplex is a vertex, a 1-simplex is an edge, a 2-simplex is a triangle,  a 3-simplex is a tetrahedon, and so on; see Figure \ref{fig:simplices}. These simplices satisfy two properties.  First, for every set $\sigma$ in $S$, every non-empty subset $\tau \subseteq \sigma$ also belongs in $S$. For instance, if tetrahedron $abcd$ is in $S$, then the triangles $abc$, $abd$, $acd$, $bcd$, the edges $ab$, $ac$, $ab$ and the vertices $a$, $b$, $c$, $d$ are also in $S$.  Second, two $k$-simplices are either disjoint or they intersect in a lower dimensional simplex.

\begin{figure}[h]
\begin{center}
\includegraphics[width=0.9\textwidth]{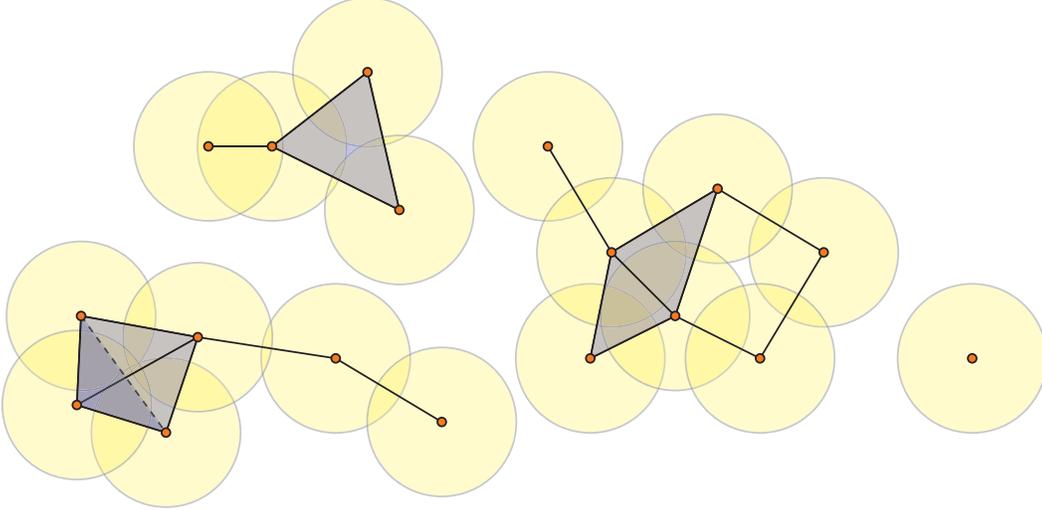} 
\end{center}
\caption{\label{fig:VR}
{\bf Example of a Vietoris-Rips complex.}  The 18 points are  0-simplices. Two 0-simplices form a 1-simplex (an edge) if their $\proximity/2$-neighborhoods (yellow circles) intersect.  Three vertices form a 2-simplex (a triangle)  if they are pairwise connected by edges. Four vertices form a 3-simplex (a tetrahedron) if they are pairwise connected by edges.}
\hrule
\end{figure}

To form $k$-simplices, we use the \emph{Vietoris-Rips complex}, sometimes simply called the Rips complex. To build this complex, one first defines a distance metric which can be realized as a symmetric $N \times N$ matrix of pairwise distances between points. For each $\proximity>0$, called the \emph{proximity parameter}, we construct a simplicial complex $S_\proximity$ in the following way. In $S_\proximity$, every collection of $k+1$ data points is a $k$-simplex if the pairwise distance between points is less than $\proximity$.  Thus, the 0-simplices are the data points themselves.  A 1-simplex (an edge) is formed whenever two points are within $\proximity$ of one another.  A 2-simplex (a triangle) is formed whenever three points are pairwise within $\proximity$ of one another; this occurs when there is a 3-cycle in the underlying graph formed by the vertices and edges in $S_\proximity$  (this graph is sometimes called the \emph{1-skeleton} of $S_\proximity$).  A 3-simplex (a tetrahedron) is formed whenever four points are pairwise within $\proximity$ of one another. 
See Figure \ref{fig:VR}, in which the yellow circles represent $\proximity/2$ balls so that two vertices are connected by an edge if their $\proximity/2$-balls intersect.

For purposes of homology, it is necessary to impose an orientation on the vertices of each $k$-simplex.  A $k$-simplex $[v_0,v_1, \ldots, v_k]$ of $k+1$ data points is ordered such that if elements are permuted with an odd permutation, the order is negated.  Thus, $[v_0,\ldots,v_i,\ldots,v_j,\ldots,v_k]=-[v_0,\ldots,v_j,\ldots,v_i,\ldots,v_l]$.  The simplices  in Figure~Œ\ref{fig:simplices} are illustrated with a given orientation.

There are other methods one could use to form a simplicial complex.  The Rips complex is the flag or clique complex that is the maximal simplicial complex built from the  underlying graph.   Another commonly used method is the \v{C}ech complex, where $k$-simplices are formed if the $\proximity/2$-balls around $k+1$ vertices have a nonempty intersection.  We use the Rips complex since it is more computationally tractable than the \v{C}ech complex; see the discussion in Section 1.3 of \cite{Ghr2008}.  The witness complex is another method to form a simplicial complex. It is particularly useful with large data sets as it subsamples the data in a way that uses information from the entire data set; see \cite{DeSCar2004}.

\subsection*{Homology}

Homology is a way to uncover $k$-dimensional  ``holes" in a simplicial complex.  This requires imposing an algebraic structure on the simplicial complex $S_\proximity$.  For each $k\geq 0$, we create an abstract vector space $\C_k$ with basis consisting of the set of $k$-simplices in $S_\proximity$, so that the dimension of $\C_k$ equals the number of $k$-simplices. The elements of $\C_k$ are called \emph{$k$-chains}.  In practical computations, the coefficients of this vector space come from a finite field  $\ZZ_p$ (integers modulo $p$) for a small prime $p$.  These vector spaces consist of all formal linear combinations $c=  \sum_i a_i \sigma_i$, where $a_i \in \ZZ_p$ and the sum is over all $k$-simplices $\sigma_i$ in $S_\proximity$. 

To compute homology, one must be able to describe the boundary of a $k$-simplex algebraically. The \emph{boundary} of a $k$-simplex $\sigma$ is the union of the $(k-1)$-subsimplices $\tau \subseteq \sigma$.  For each $k \ge 1$, the \emph{boundary map}
$
\partial_k: \C_k \to \C_{k-1}
$
is the linear transformation defined on a $k$-simplex $\sigma = [v_0,v_1, \ldots, v_k]$  by
\begin{equation}
\partial_k([v_0,v_1, \ldots, v_k]) = \sum_{i=0}^k (-1)^i [v_0, \ldots, \hat{v_i}, \ldots, v_k],
\end{equation}
where $[v_0, \ldots, \hat{v_i}, \ldots, v_k]$ is the $(k-1)$-simplex obtained from $[v_0, \ldots, v_k]$ by removing the vertex $v_i$.    For example,
$$
\partial_1([v_0,v_1]) = [v_1] - [v_0]  \qquad\hbox{and}\qquad \partial_2([v_0,v_1,v_2]) = [v_1,v_2] - [v_0,v_2] + [v_0,v_1].
$$
Observe that for a $k$-simplex $\sigma$, $\partial_k(\sigma)$ is an algebraic representation of its boundary. For instance, in the second example above, $[v_0,v_1,v_2]$ represents a triangle and $[v_1,v_2] - [v_0,v_2] + [v_0,v_1]$ are the oriented edges that form its boundary.

Boundary operators connect the vector spaces $\C_k$ into a \emph{chain complex},
\begin{equation}
\cdots\  {\longrightarrow}\   \C_{k+1} \stackrel{\partial_{k+1}} {\longrightarrow} \C_k  \stackrel{\partial_{k}}{\longrightarrow}  \C_{k-1}   {\longrightarrow}  \cdots 
{\longrightarrow}\ \C_2 \stackrel{\partial_{2}}{\longrightarrow}\  \C_1 \stackrel{\partial_{1}}{\longrightarrow}  \C_0 \stackrel{\partial_{0}}{\longrightarrow} 0.
\end{equation}
Consider the following two subspaces of $\C_k$, which are determined by the kernel and the image of the boundary operators,
\begin{equation}
\begin{array}{ll}
\text{$k$-cycles:} & \Z_k := \ker\left(\partial_k: \C_k \to \C_{k-1}\right), \\
\text{$k$-boundaries:} & \B_k := \im\left(\partial_{k+1}: \C_{k+1} \to \C_{k}\right). \\
\end{array}
\end{equation}
The boundary operator satisfies the following fundamental property: $\partial_{k} \circ \partial_{k+1} = 0$.  That is, ``a boundary has no boundary".  For example, 
$$\partial_1 \circ \partial_2([v_0,v_1,v_2]) = \partial_1([v_1,v_2]) -  \partial_1([v_0,v_2]) +  \partial_1([v_0,v_1]) = [v_2]-[v_1] -[v_2] + [v_0] + [v_1] - [v_0] = 0.$$  It then follows that $\B_k$ is a subspace of $\Z_k$.  Thus, $\C_k$ is the vector space of all $k$-chains in the simplicial complex $S_\proximity$, $\Z_k$ is the subspace of $\C_k$ consisting of  $k$-chains that are also $k$-cycles, and $\B_k$ is the subspace of $\Z_k$ consisting of $k$-cycles that are also $k$-boundaries.  

\begin{figure}[t]
\begin{center}
\includegraphics[width=0.5\textwidth]{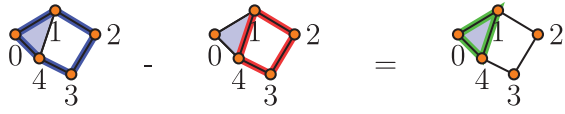}
\end{center}
\caption{\label{fig:EquivalentCycles}
{\bf Example of homologous cycles.} The blue 1-cycle and the red 1-cycle are homologous (equivalent), because their difference is the boundary of a triangle, shown in green; see text for a detailed explanation.}
\hrule
\end{figure}

The goal of homology is to ``discard"  cycles that are also boundaries.   To this end, we put an equivalence relation on $\Z_k$ as follows. Two cycles $z_1, z_2 \in \Z_k$ are \emph{homologous} (equivalent), written $z_1 \sim z_2$, if they differ by a boundary, \emph{i.e.}, $z_1 - z_2 \in \B_k$.  Consider the example in Figure~\ref{fig:EquivalentCycles}. The blue 1-chain $b = [v_0,v_1] + [v_1,v_2] + [v_2,v_3] + [v_3,v_4] + [v_4,v_0] $ and the red 1-chain $r = [v_1,v_2] +[v_2,v_3] + [v_3,v_4] + [v_4,v_1]$ are cycles because $\partial_1(b) = \partial_1(r)=0$. These cycles are homologous because their difference  is a (green) boundary, $g = b - r =  [v_0,v_1]  + [v_4,v_0] - [v_4,v_1] =   [v_1,v_4]  - [v_0,v_4]  +  [v_0,v_1] =\partial [v_0,v_1,v_4]$.

The equivalence relation $\sim$ defined above partitions the $k$-cycles $\Z_k$ into a union of disjoint subsets, called \emph{homology classes}.  We let $[z]$ denote the homology class of $z \in \Z_k$ and define the $k$th homology of $S_\proximity$ as the set of homology classes
\begin{equation}
\H_k := \left\{\ [z] \  \big\vert\  z\ \in \Z_k\ \right\}.
\end{equation}
Algebraically, $\H_k = \Z_k/\B_k$, a quotient of vector spaces. The $k$th Betti number, $\betti_k$, is defined as the dimension
\begin{equation}
\betti_k=\text{dim}(\H_k)=\text{dim}(\Z_k)-\text{dim}(\B_k).
\end{equation}
In terms of boundary operators, $\betti_k=[n_k - \text{rank}(\partial_k)]-\text{rank}(\partial_{k+1})$, where $n_k$ is the dimension of the vector space $\C_k$.

In terms of the topological characteristics one might hope to measure, $\betti_k$ equals the number of independent holes of dimension $k$, and this is the key point for our analysis later in this paper. For instance, $\betti_0$ is the number of connected components, $\betti_1$ is the number of topological circles, $\betti_2$ is the number of trapped volumes, and so on.  The topology of a simplicial complex may be described by the sequence of Betti numbers, $\betti=(\betti_0, \betti_1,\betti_2,\ldots)$. For instance, a topological circle has $\betti=(1,1,0,\ldots)$, a topological torus has $\betti=(1,2,1,0,\ldots)$, and a topological sphere has $\betti=(1,0,1,0,\ldots)$.  Betti numbers are a topological invariant, meaning that topologically equivalent spaces have the same Betti number. See, \emph{e.g.}, \cite{Hat2002,Cro2006} for these and other examples.

\subsection*{Persistence}

\begin{figure}[t]
\begin{center}
\includegraphics[width=\textwidth]{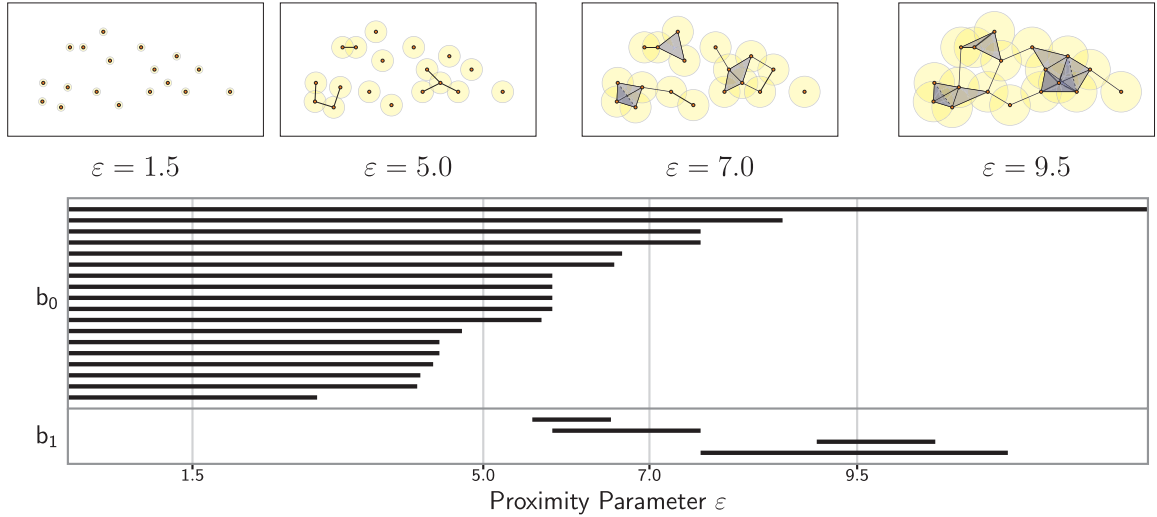}
\end{center}
\vspace{-0.3in}
\caption{\label{fig:pedExample}
{\bf Example of the topological barcode of a Vietoris-Rips complex.} The top four figures display the simplicial complex of 18 points for  different values of the proximity parameter $\proximity$. The vertical lines in the barcode correspond to these four levels of $\proximity$.  The number of horizontal bars intersecting each line give the values of $\betti(\proximity) = (\betti_0(\proximity),\betti_1(\proximity))$.  For the parameters selected, $\betti(1.5)=(18,0)$, $\betti(5.0)=(11,0),$ $\betti(7.0)= (4,1),$ and $\betti(9.5)=(1,2)$. See text for further discussion.}
\hrule
\end{figure}

Given a collection of $N$ data points, the resulting Rips complex and its homology are highly dependent on the choice of proximity parameter $\proximity$. Figure~\ref{fig:pedExample} presents an example. The data pictured in the four small snapshots are the same as Figure~\ref{fig:VR}, but different values of $\proximity$ are chosen for forming connections.  Distinct simplicial complexes result. For small values of $\proximity$, the simplicial complex consists of isolated vertices. At the largest value of $\proximity$ shown, the entire data set is a single connected component. As $\proximity$ changes, other topological events occur which we will describe momentarily. A natural question is what is the optimal $\proximity$ to use for any data set. This can hardly be selected without \emph{a priori} knowledge of the underlying space.  

To reconcile this ambiguity, one exploits the fact that as $\proximity$ grows so do the Rips complexes, giving an inclusion of complexes for small $\proximity$ into those for larger values. That is, if $\proximity_1 \leq \proximity_2 \leq \cdots \leq \proximity_M$, then we have an inclusion of simplicial complexes
\begin{equation}
S_{\proximity_1} \subseteq S_{\proximity_2} \subseteq \cdots \subseteq S_{\proximity_{M-1}} \subseteq S_{\proximity_M}.
\end{equation}
This sequence is called a \emph{filtration}.  ``Persistent'' homology, then, tracks topological features which persist across a range of values of $\proximity$.  Those features which persist over a large range are considered signals of the underlying topology, while the short lived features are taken to be noise inherent in approximating a topological space with a finite sample. Foundational papers on this methodology include \cite{EdeHar2008,Ghr2008,Car2009}.

A convenient way to visualize persistent homology is through a graphical representation called a \emph{barcode}. There is a distinct barcode for each homology space $\H_k$ from which we infer the Betti number $\betti_k$. As an example, see Figure~\ref{fig:pedExample}. The horizontal axis corresponds to the proximity parameter $\proximity$, and the vertical axis is an (arbitrary) ordering of the homology generators, \emph{i.e.}, the distinct homology classes of dimension $k$.  Each homology class is visualized by a bar that persists for a given range of $\proximity$.  Its leftmost endpoint is at the $\proximity$ value at which the homology class forms, and its rightmost endpoint is the $\proximity$ value at which it disappears. At any given $\proximity$, the Betti number $\betti_k(\proximity)$ is the number of bars that intersect the vertical line through $\proximity$.  Those bars which persist over longer intervals generally correspond to real topological features, whereas short bars are considered noise. In the figure, we see the following sequence of Betti numbers. For $\proximity = 1.5$, $\betti = (18,0,\ldots)$ because there are no connections amongst the 18 vertices. For $\proximity = 5$, $\betti = (11,0,\ldots)$, reflecting the fact that some vertices have joined into larger connected components. For $\proximity = 7$, $\betti=(4,1,0,\ldots)$, reflecting even further joining of components as well as the formation of one topological circle. Finally, for $\proximity = 9.5$, $\betti=(1,2,0,\ldots)$, meaning that all of the data has joined into one connected component that contains two topological circles.

Another method of displaying homological information is through a \emph{persistence diagram}.  Each bar in the barcode is represented in the Cartesian plane with the horizontal and vertical axes encoding the leftmost and rightmost $\proximity$ values of the bar. Points near the diagonal are inferred to be noise while points further from the diagonal are considered topological signal. Figure~\ref{fig:vicsekBarcodes}(D) shows the persistence diagram corresponding to two barcodes in panels (B) and (C). This data is generated by a biological aggregation model which we discuss later.

Because our data sets are obtained from numerical simulation, they are noiseless. Still, one might wonder whether small perturbations of data would impact the topological features that are measured. For biological aggregations, this question would be especially relevant in analyzing experimental data, for which measurement error might introduce noise. As shown in \cite{CohEdeHar2007}, small perturbations of data result in small perturbations of persistence diagrams, indicating stability of topological features.

\section*{Computational Methods and Data Visualization}

In the next two sections, we will present topological analyses of simulation output from the two aggregation models of \cite{VicCziBen1995} and \cite{DOrChuBer2006}. We perform the simulations in M{\sc atlab}. For the Vicsek model \cite{VicCziBen1995}, as we describe in more detail in the next section, the simulation output consists of the two dimensional position and the angular heading of each agent in a group of interacting agents. The physical domain is a square with sides of length $\ell$ and periodic boundary conditions. Heading is defined on $[0,2\pi)$. To avoid issues arising from disparities between the position coordinates and the heading coordinate, we rescale position coordinates, multiplying them by $2\pi/\ell$. For the D'Orsogna model \cite{DOrChuBer2006}, the simulation output consists of the two dimensional position and two dimensional velocity of each agent in an interacting group. The spatial domain is an unbounded plane and velocity is (theoretically) unbounded, so we perform no rescaling of coordinates.

The computational complexity of computing $\betti_k$ depends on the number of $k$-simplices. For $n$ data points, the number of $k$-simplices is at most $\binom{n}{k+1} = \Order(n^{k+1})$; of course, the actual number depends on the proximity parameter $\proximity$ and the configuration of the data. Once the simplicial complex is constructed, computing homology over a field reduces to methods in linear algebra. The boundary operator $\partial_k:\C_k \to \C_{k-1}$ is the linear transformation realized as an integer matrix with entries $\{-1,0,1\}$ over the basis elements of each vector space.  The null space of this matrix corresponds to $\Z_k$, and the range space corresponds to $\B_{k-1}$. The computational algorithm uses Gaussian elimination over a finite field, which is at worst case $\Order(m^3)$, where $m$ is the actual number of $k$-simplices. One needs efficient algorithms when computing over large data sets and recording homological information over a range of proximity parameter $\proximity$. See \cite{ZomCar2005,KacMisMro2004, EdeLetZom2002, BauKerRei2014} for implementations and algorithms used in computing persistent homology. The development of faster algorithms is an active area of investigation.

To extract topological information, we process simulation data in the statistical computing environment \textsf{R}. Each time step of the simulation consists of a static point cloud of data in position-heading or position-velocity space. We use the \texttt{phom} package \cite{Tau2011} to construct the Rips complex and calculate the topological barcodes of the point cloud. More specifically, we calculate the first two Betti numbers $\betti_0(\proximity)$ and $\betti_1(\proximity)$ where the proximity parameter $\proximity$ takes on discrete but closely-spaced values. Calculating $\betti_k(\proximity)$ for $k \geq 2$ is computationally costly, and we do this only for selected simulation snapshots.

Because we have a series of simulation time steps, we introduce a visualization that captures homological persistence over both scale $\proximity$ and time $t$. That is, we now imagine the $k$th Betti number as $\betti_k(\proximity,t)$, a function of the proximity parameter $\proximity>0$ and  simulation time $t\ge 0$. A natural way to display $\betti_k(\proximity,t)$ is as a contour diagram, which we refer to as a Contour Realization Of Computed $k$-dimensional hole Evolution in the Rips complex (CROCKER). To facilitate visual interpretation of the contour diagram, we make two simplifications. First, we do not include every time step, but rather only every $j$th time step, where $j$ is chosen to preserve at least several hundred snapshots per simulation. We found that this downsampling did not noticeably alter the appearance of our CROCKER plots. Second, focusing only on the most coarse and coherent topological structures, we only plot level curves for when $\betti_k(\proximity,t) < 5$. All of the topological data for which $\betti_k(\proximity,t) \geq 5$ is lumped together in the contour diagram; that is, we do not draw distinct contours for $\betti_k(\proximity,t) \geq 5$. For the models we study, these regions of the contour diagram typically represent many topological structures that do not persist over scales.

As an example, consider Figure~\ref{fig:vicsek1bTimeSeries}. Panels (B) and (C) show contour diagrams of $\betti_{0}(\proximity,t)$ and $\betti_{1}(\proximity,t)$ for a particular simulation of the Vicsek model (described and analyzed later). In these plots, simulation time $t$ appears along the horizontal axis and the topological proximity parameter $\proximity$ appears on the vertical axis. Focus first on panel (B), which shows $\betti_0(\proximity,t)$. Below the purple (lowest) contour, $\betti_0(\proximity,t) \geq 5$, and thus in this region, the point cloud has many connected components; we interpret these as noise. Above the yellow (top) contour, $\betti_0(\proximity,t) = 1$, demonstrating that at large enough $\proximity$, the entire point cloud of data joins together into one connected component. Less trivially, there are regions between the intermediate contours that persist over scale and time. For example, centered near $t = 2000$ there is a somewhat triangular region between the yellow and green contours. In this region, $\betti_0(\proximity,t) = 2$, showing a strong, persistent signal of two connected components in the data. Panel (C) is similar, but shows $\betti_1(\proximity,t)$. Towards the bottom of the diagram, there is an oblong region enclosed by a purple contour. In this region, $\betti_1(\proximity,t) \geq 5$, which we again interpret as topological noise. There are two large regions enclosed by a red contour in which $\betti_1(\proximity,t) = 0$, indicating an absence of topological circles. However, there is also a region between red and yellow contours in which $\betti_1(\proximity,t) = 1$, showing a strong signal of a persistent topological circle.

In summary, large regions in the contour diagram (excluding $\betti_k \geq 5$) represent topological features that persist over scale $\proximity$ and simulation time $t$. When interpreting the contour diagrams, it is important to remember that the function $\betti_k(\proximity,t)$ inherently takes on only nonnegative integer values.

\section*{Analysis of the Vicsek Model}
\label{sec:vicsek}

Using persistent homology, we now analyze data generated by aggregation models. One of the most referenced aggregation models is that of Vicsek and collaborators \cite{VicCziBen1995}, cited thousands of times as of the writing of this manuscript. A complete discussion of results related to this model is beyond our present scope. The review paper \cite{VicZaf2012} provides a broad look, and mentions some of the systems that have been described with Vicsek-like models, including cells, bacteria, insects, fish, and birds.

The Vicsek model is a dynamical system in discrete time and continuous space that describes the motion of interacting point particles in a square with periodic boundary conditions. The model appears in the literature written in different forms; we write it as
\begin{subequations}
\label{eq:vicsekModel}
\begin{eqnarray}
\theta_i(t+\Delta t) & = & \frac{1}{N} \left(\sum_{|\vec{x}_i - \vec{x}_j| \leq R} \theta_j(t) \right) + U(-\eta/2,\eta/2),\label{eq:vicsekModel1}\\
\vec{v}_i(t+\Delta t) & = & v_0 \bigl(\cos \theta_i(t+\Delta t),\sin \theta_i(t+\Delta t) \bigr),\\
\vec{x}_i(t+\Delta t) & = & \vec{x}_i(t) + \vec{v}_i(t+\Delta t)\, \Delta t.
\end{eqnarray}
\end{subequations}
Here, $\vec{x}_i(t) \in \mathbb{R}^2$ is the position of particle $i=1,
\ldots,N$ at time $t$, and $\vec{v}_i$ is velocity. We refer to the angle of the velocity vector $\vec{v}_i$ as $\theta_i$, the heading. Additionally, $v_0$ is a constant, $U$ is a uniform random variable on the specified interval, and $\Delta t$ is the time step.

For clarity, let us re-state the model in prose. To update the model, each particle must be given a new heading. This heading is the average of the previous headings of all other particles within a radius $R$, plus some added noise parameterized by $\eta$. With this new heading determined, each particle  moves a fixed distance $v_0\Delta t$, thus completing the time step. The model is posed on a square with sides of length $\ell$ and periodic boundary conditions. For all simulations performed, the initial particle positions and headings are random.

The parameters in the model are the number of particles $N$, the particle interaction radius $R$, the noise $\eta$, the fixed particle speed $v_0$, the box size $\ell$, and the time step $\Delta t$. One may nondimensionalize the problem to reduce the number of free parameters. We adopt the standard convention $R=1$ and $\Delta t = 1$. This leaves the remaining parameters $N$, $\eta$, $v_0$, and $\ell$. Some studies refer to three effective parameters: $\eta$, $v_0$, and $\rho = N/\ell^2$, a particle density.

Two preliminary matters will build understanding prior to a discussion of results. First, we analyze the topology of an initial condition. Second, we discuss the classic order parameter used in the physics literature to characterize the global dynamics of the system.

\begin{figure}[ph!]
\begin{center}
\includegraphics[width=0.9\textwidth]{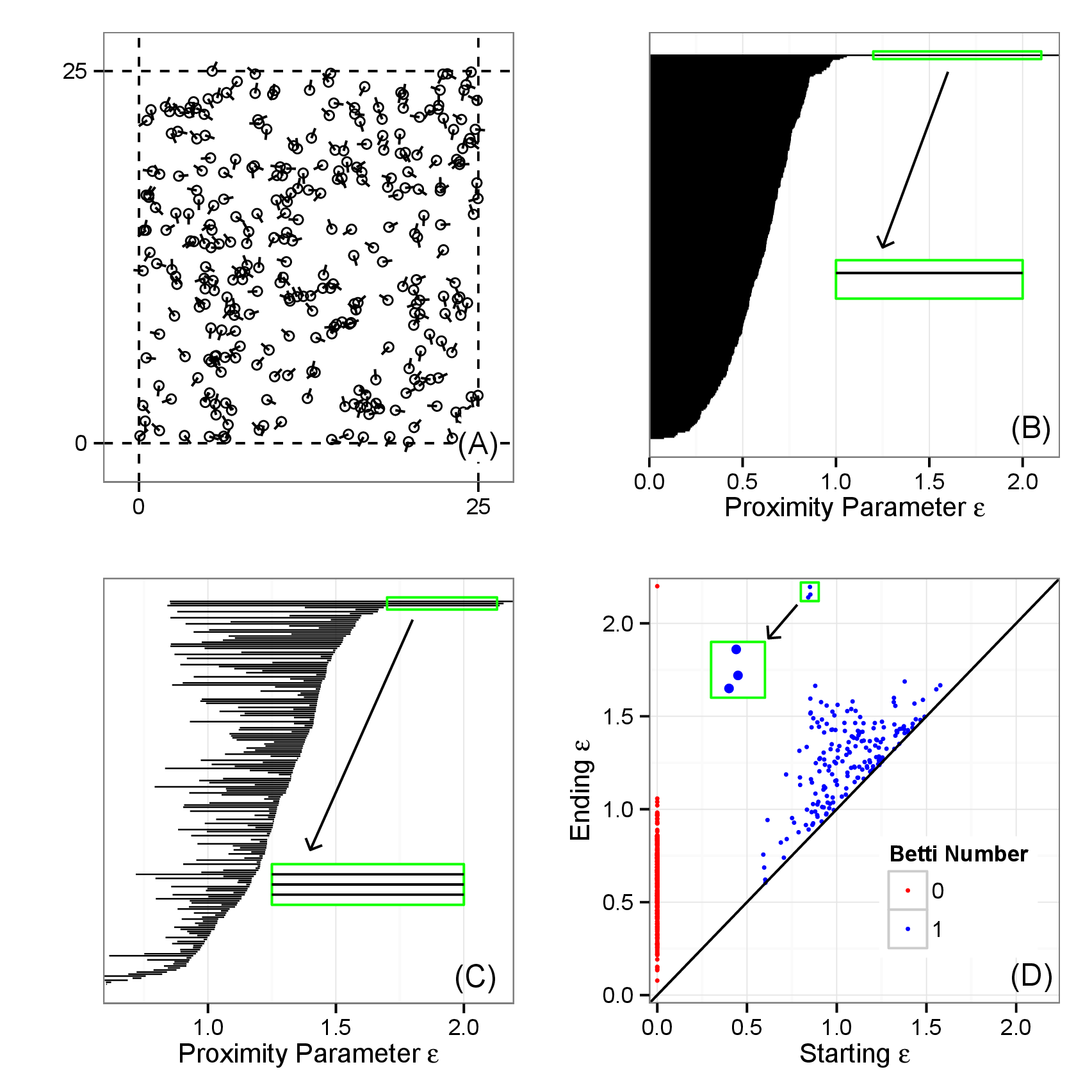}
\end{center}
\caption{\label{fig:vicsekBarcodes}
{\bf A random initial condition used to simulate the Vicsek model (\ref{eq:vicsekModel}) and topological analysis of this initial state.} (A) Random initial positions $(x,y)$ and headings $\theta$ of $N=300$ particles in a square of size $\ell=25$ with periodic boundary conditions. The underlying space in which the data lives is a three-torus $\mathbb{T}^3$ which has Betti numbers $\betti=(1,3,3,1,0,\ldots)$. (B) Barcode for Betti number $\betti_0(\proximity,0)$, showing topological connected components. The zoomed box shows a single persistent bar, corresponding to the entire ensemble of particles. (C) Barcode for Betti number $\betti_1(\proximity,0)$, showing topological circles. The zoomed box shows three persistent bars, representing the three circles comprising the three-torus. (D) Persistence plot, which displays the information in (B) and (C) by encoding each bar's starting and ending value of $\proximity$ as a point in the Cartesian plane. Red points show $\betti_0$ and blue points show $\betti_1$. The zoomed box shows the three points representing the three persistent topological circles of the random initial condition in (A).}
\end{figure}

Figure~\ref{fig:vicsekBarcodes}(A) shows a random initial condition for $N=300$ particles and a square with sides of length $\ell = 25$. This gives rise to a cloud of $300$ points whose coordinates are $(x,y,\theta)$. However, since the simulation domain is periodic and since $\theta$ is an angle, this space is not $\mathbb{R}^3$ but rather $S^1 \times S^1 \times S^1 = \mathbb{T}^3$, the three-torus, which has Betti numbers $\betti=(1,3,3,1,0,\ldots)$. Panel (B) shows the topological barcode for $\betti_0(\proximity,0)$, which has one persistent bar. This topological signature indicates no clusters other than the trivial connected component formed by the entire point cloud on the longest scales. Panel (C) shows the barcode for $\betti_1(\proximity,0)$. There are three persistent bars (which terminate for larger values of $\proximity$ not shown) representing three topological circles. This signature captures the fact that the point cloud is well-spread over the three-torus due to the randomness of the initial condition. We have also calculated the barcode (not shown) for $\betti_2(\proximity,0)$, which displays three persistent bars, representing the three trapped volumes of the three-torus. Panel (D) shows the persistence diagram, which combines the information captured in (B) and (C). The particular initial condition we have analyzed is used to seed our first simulation below. The initial conditions of the other two simulations that we will perform are topologically equivalent. Below, we will see that key dynamics involve the formation of nontrivial topological connected components and/or the destruction of topological circles. 

\begin{figure}[ph!]
\begin{center}
\includegraphics{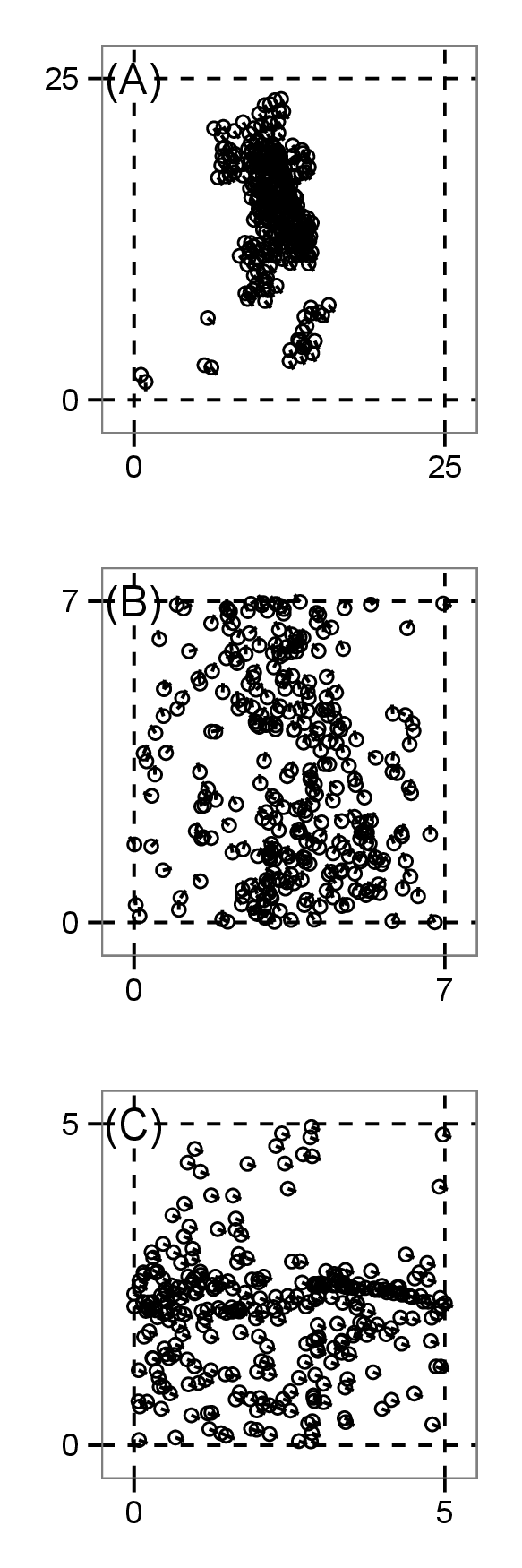}
\end{center}
\caption{\label{fig:vicsekStates}
{\bf Simulation snapshots of the Vicsek model (\ref{eq:vicsekModel}).} These simulations are analogous to Figure 1 in \cite{VicCziBen1995}. Circles indicate particle positions and line segments represent heading. For all simulations, $N = 300$ particles, the particle speed is $v_0=0.03$, and the initial state consists of uniform random positions and headings. We vary box size $\ell$ and noise $\eta$. Dotted lines indicate the bounds of the periodic domain. (A) Groups moving in different directions with $\ell=25$, $\eta = 0.1$, $t=3000$. (B) Random movement with some correlation with $\ell=7$, $\eta=2$, $t=600$. (C) Highly polarized motion with $\ell=5$, $\eta=0.1$, $t=300$.}
\end{figure}

As discussed in the introduction, a traditional approach is to characterize global behavior via an order parameter. For (\ref{eq:vicsekModel}), the order parameter most often studied is the normalized average velocity of the group,
\begin{equation}
\varphi(t) = \frac{1}{Nv_0} \left| \sum_{i=1}^{N}\vec{v}_i(t)\right|.
\end{equation}
The order parameter $0 \leq \varphi(t) \leq 1$ measures global polarization. For a group of particles moving in approximately the same direction, $\varphi(t)$ will be near one. If particle headings are spread out randomly, $\varphi(t)$ will be near zero. As a simple additional example, two groups of particles that are highly polarized within each group but are traveling in opposite directions will also have $\varphi(t) \approx 0$. This simple example sheds some light on the limited information the order parameter carries.

The model (\ref{eq:vicsekModel}) can display three qualitatively different global behaviors, depending on parameters. We visualize snapshots of these states in Figure~\ref{fig:vicsekStates}, which is analogous to Figure~1 of Vicsek's original paper \cite{VicCziBen1995}. When noise $\eta$ is small and particle density $\rho$ is small, the system tends to form clusters each of which moves in a different direction, as in (A). For higher $\eta$ and $\rho$, particles are somewhat correlated but still move randomly, as in (B). Finally, for large $\rho$ but small $\eta$, the motion becomes polarized, that is, all particles travel in the same direction, as in (C). For the simulations represented by these three panels, we will now view the global behavior through the lens of the order parameter $\varphi(t)$ and through a topological lens. The topological analysis gives rich information which is detected neither by the order parameter nor, as we will show, by the eye.

\subsection*{Vicsek Simulation \#1}

\begin{figure}[ph!]
\begin{center}
\includegraphics[width=0.9\textwidth]{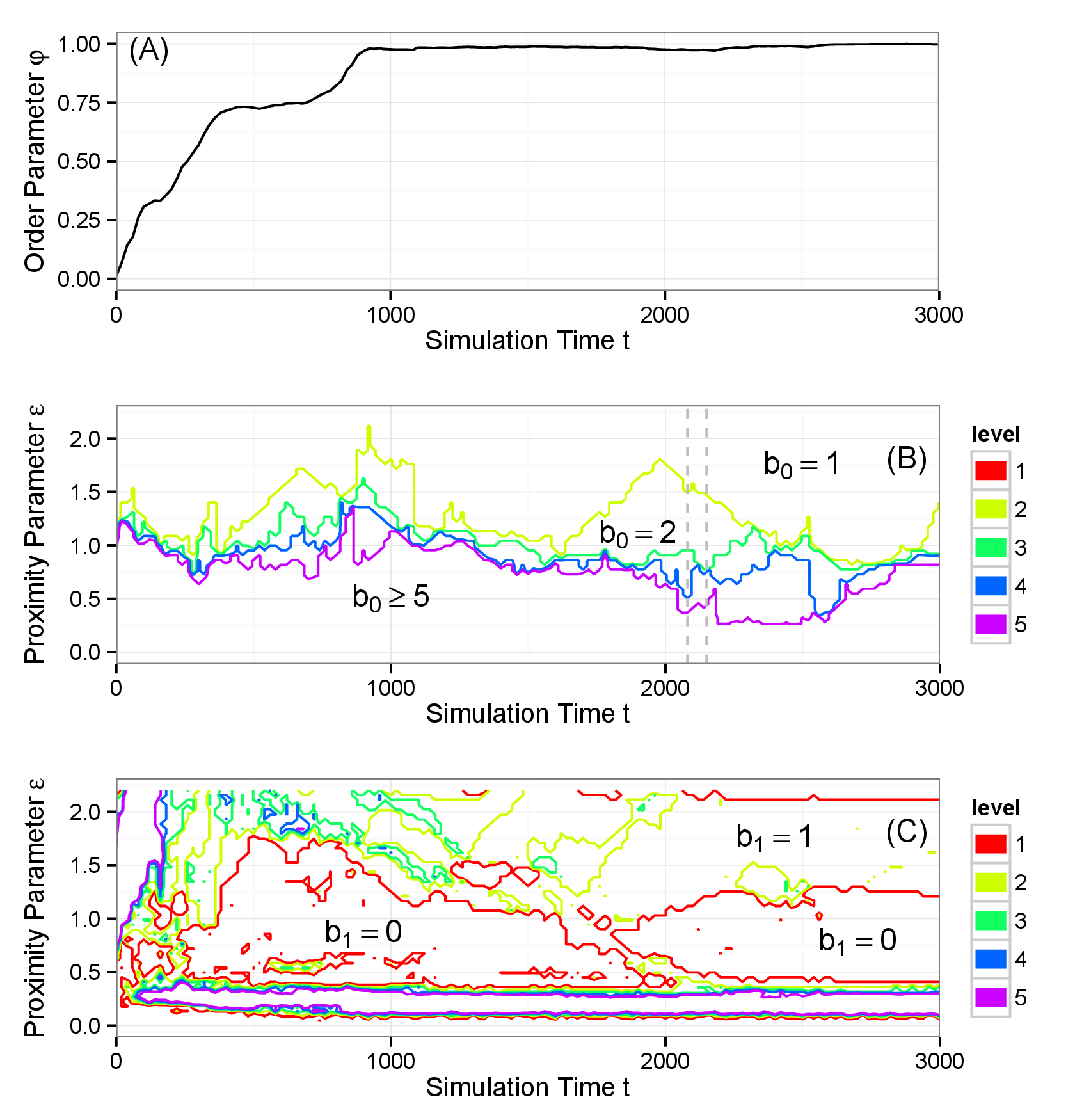}
\end{center}
\caption{\label{fig:vicsek1bTimeSeries}
{\bf Aggregate behavior of the Vicsek model, simulation \#1.} The simulation that generated these data was seeded with the initial condition in Figure~\ref{fig:vicsekBarcodes}(A) and a typical snapshot is shown in Figure~\ref{fig:vicsekStates}(A). (A)  Normalized average velocity order parameter $\varphi(t)$. (B) Contour plot of Betti number $\betti_0(\proximity,t)$. (C) Contour plot of Betti number $\betti_1(\proximity,t)$. The topological analysis reveals dynamics not captured by the order parameter, namely cluster formation and the loss of topological circles consistent with particles aligning and covering only one dimension of the periodic simulation domain. See text for a more comprehensive analysis.}
\end{figure}

Figure~\ref{fig:vicsek1bTimeSeries} displays the analysis of our first set of simulation results, the same simulation that began with the initial condition in Figure~\ref{fig:vicsekBarcodes}(A) and produced the state in Figure~\ref{fig:vicsekStates}(A). Here, $N=300$, $\ell=25$, and $\eta=0.1$. The order parameter $\varphi(t)$ increases steeply (with two small plateaus) before leveling off to a value very close to one, signaling a high degree of alignment. After time $t=1000$, $\varphi(t)$ varies little. 

Figure~\ref{fig:vicsek1bTimeSeries}(B) shows $\betti_0(\proximity,t)$, measuring the number of connected components on the three dimensional torus defined by position and heading. Recall that, per our previous discussion, we have displayed only contours for levels five and below. In the large region below the purple (bottom) contour, $\betti_0(\proximity,t) \geq 5$. This region represents connected components existing only over small ranges of $\proximity$; we interpret these as noise. In the region above the yellow (top) contour,  $\betti_0(\proximity,t) = 1$, and all the data forms a single connected component for large $\proximity$. The spaces between the other contours reveal clusters that persist over scale and simulation time. For example, there is a triangular region between the green and yellow contours around $t \approx 2000$, indicating a strong signal of two clusters. 

\begin{figure}[t!]
\begin{center}
\includegraphics[width=0.9\textwidth]{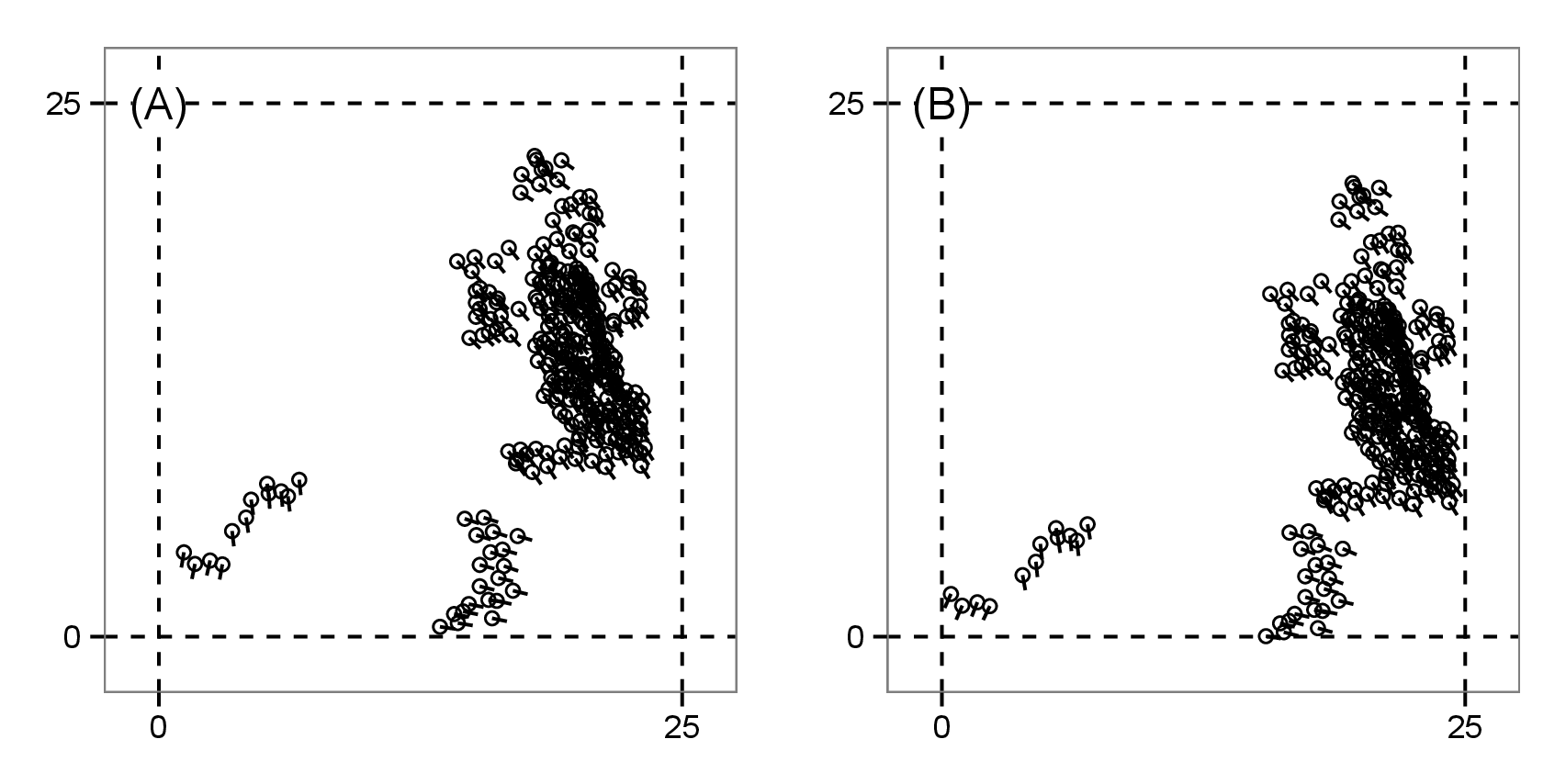}
\end{center}
\caption{\label{fig:vicsek1bLaterSnapshots}
{\bf Snapshots of the Vicsek model, simulation \#1.} These states correspond to the dashed vertical bars in Figure~\ref{fig:vicsek1bTimeSeries}(B). (A) Time $t=2080$. (B) Time $t=2150$. The topological signature in Figure \ref{fig:vicsek1bTimeSeries}(B) picks up subtle differences between these states. For panel (A) here, there are ranges of the persistence parameter $\proximity$ over which one observes four, three, and two connected components before coalescing into one. For (B), the transition from four connected components to two happens over a much smaller range of $\proximity$ because the two larger clusters on the right merge into one on approximately the same spatial scale that the two smaller clusters on the left do. The topological differences between (A) and (B) are not readily visible to the eye in the snapshots, nor are they reflected in the order parameter $\varphi(t)$ in Figure \ref{fig:vicsek1bTimeSeries}(A).}
\hrule
\end{figure}

Further analyzing this region, consider the times marked by the two dashed gray bars, namely $t=2080$ and $t=2150$. While the order parameter $\varphi(t)$ (by design) does not detect clusters, at $t=2080$, the spaces between the contours reveal successive ranges of $\proximity$ over which there are four, three, two, and one connected components in the data, with $\betti_0(\proximity,2080)=2$ persistent over the largest range of $\proximity$, as previously discussed. The corresponding simulation snapshot appears in Figure~\ref{fig:vicsek1bLaterSnapshots}(A). The transition from four to three connected components corresponds to the merging of two small, slightly misaligned groups in the lower left of Figure~\ref{fig:vicsek1bLaterSnapshots}(A) as $\proximity$ increases. The transition from three to two connected components corresponds to the merging of the two larger groups and so on. The situation for $t=2150$ is similar. However, the range of $\proximity$ over which one counts three connected components has shrunk drastically. The simulation snapshot in Figure~\ref{fig:vicsek1bLaterSnapshots}(B) suggests why. In Figure~\ref{fig:vicsek1bLaterSnapshots}(A), the distance between the two leftmost groups approximately equals the distance between the two rightmost groups. Hence, there does not exist a large range of $\proximity$ over which there are three clusters. It is important to remember that the data has three coordinates $(x,y,\theta)$, but in the discussion above, the key dynamics are in $(x,y)$, consistent with the fact that $\varphi(t) \approx 1$.

Figure~\ref{fig:vicsek1bTimeSeries}(C) shows $\betti_1(\proximity,t)$, measuring the number of topological circles formed by the data on the three-torus. First, we note that the three topological circles present in the initial condition (described above) are lost on a very short simulation time scale (nearly immediately). In the bottom, oblong region of the graph enclosed by the purple contour, $\betti_1(\proximity,t) \geq 5$ which, as before, we interpret as noise. There are two large regions enclosed by red in which there exist no topological circles. However, for approximately $t > 1800$ and $1.25 \leq \proximity \leq 2.1$ we have $\betti_1(\proximity,t) = 1$, consistent with coverage of the data across one of the circles of the underlying three-torus. This circle is visible in Figure~\ref{fig:vicsek1bLaterSnapshots} as the vertical swath of data along the right side of each panel. The fact that there is a high degree of alignment removes one of the potential topological circles of the underlying space, and the lack of coverage horizontally across the spatial domain removes the other.

Taken together, panels (B) and (C) of Figure~\ref{fig:vicsek1bTimeSeries} show the following topology. First, $\betti_0(\proximity,t)$ reveals that a small number of clusters form persistently over time and scale. The number of clusters is variable as merging and fragmenting occur. Second, $\betti_1(\proximity,t)$ reveals that topological circles present in the initial condition are destroyed, but eventually one circle persists, consistent with coverage across one dimension of the simulation domain.

\subsection*{Vicsek Simulation \#2}

\begin{figure}[ph!]
\begin{center}
\includegraphics[width=0.9\textwidth]{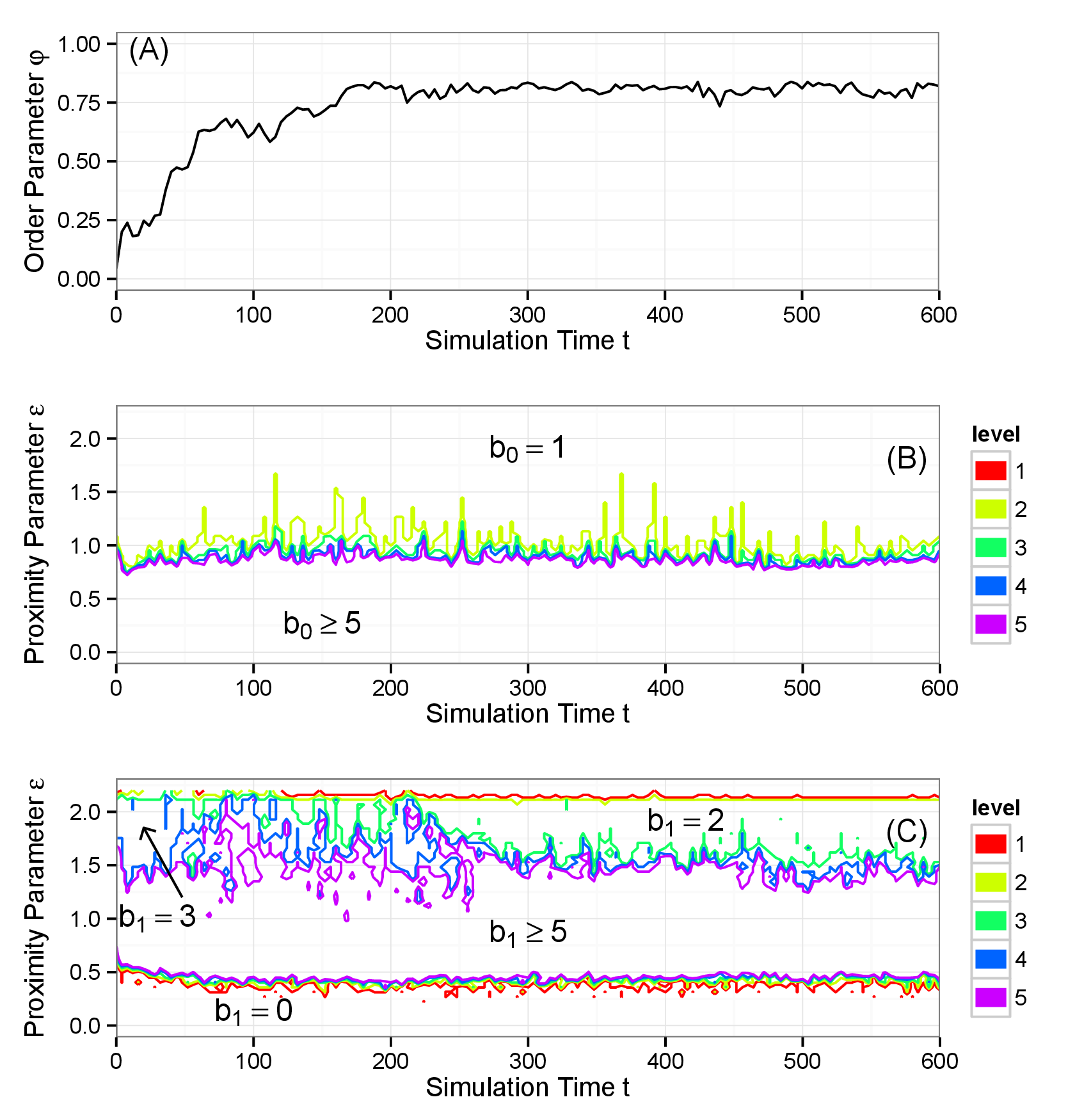}
\end{center}
\caption{\label{fig:vicsek1cTimeSeries}
{\bf Aggregate behavior of the Vicsek model, simulation \#2.} A typical snapshot is shown in Figure~\ref{fig:vicsekStates}(B). (A)  Normalized average velocity order parameter $\varphi(t)$. (B) Contour plot of Betti number $\betti_0(\proximity,t)$. (C) Contour plot of Betti number $\betti_1(\proximity,t)$. The topological analysis suggests sporadic coagulation and fragmentation of short-lived clusters, and the loss of a topological circle, consistent with particles aligning. See text for a more comprehensive analysis.}
\end{figure}

Figure~\ref{fig:vicsek1cTimeSeries} shows results from the same simulation as the snapshot in Figure~\ref{fig:vicsekStates}(B). Here, $N=300$, $\ell=7$, and $\eta=2$. In panel (A), the order parameter $\varphi(t)$ appears more noisy than in the previous simulation, and levels off to a value less than one.

Figure~\ref{fig:vicsek1cTimeSeries}(B) shows $\betti_0(\proximity,t)$, measuring the number of connected components. In the large region below the purple (bottom) contour, $\betti_0(\proximity,t) \geq 5$, which we interpret as noise. In the large region at the top, above the yellow contour, $\betti_0(\proximity, t) = 1$, indicating one connected component at the largest scales. The bottom and top regions in the graph are separated by a noisy boundary of intermediate curves. These curves indicate that at scales approximately in the range $0.75 < \proximity < 1.6$ there is sporadic coagulation and fragmentation of connected components over short simulation time scales.

Figure~\ref{fig:vicsek1cTimeSeries}(C) shows $\betti_1(\proximity,t)$, measuring the number of topological circles. At early times, there is a region in which $\betti_1(\proximity, t) = 3$, consistent with the initial condition in which agents cover the spatial domain and have low alignment, with headings spread across $[0,2\pi)$. This state persists until approximately $t \approx 50$, at which time we see noisy formation and break up of additional topological circles in the range $1.5 \leq \proximity \leq 2$. At time $t \approx 210$, a transition occurs, and we see a clear region where $\betti_1(\proximity,t) =2$, indicating two persistent topological circles. The high (albeit not complete) alignment of particles visible in $\varphi(t)$ suggests the lack of a topological circle in the heading coordinate $\theta$. We have calculated for one time step that $\betti_2(\proximity,600)=1$. This trapped volume tells us that our data has the topology of a two-torus, that is, $\betti=(1,2,1)$ for the first three Betti numbers. This topology eliminates the possibility that the two topological circles arise from holes in the data, which would have the first three Betti numbers $\betti=(1,2,0)$. We conclude that the two topological circles correspond to the periodic spatial domain. Neither dimension of this domain is favored, as suggested by the fact that the two topological circles close at the same value of $\proximity$, demonstrated by the yellow and red contours at the top of the plot being nearly coincident.

\subsection*{Vicsek Simulation \#3}

\begin{figure}[p!]
\begin{center}
\includegraphics[width=0.9\textwidth]{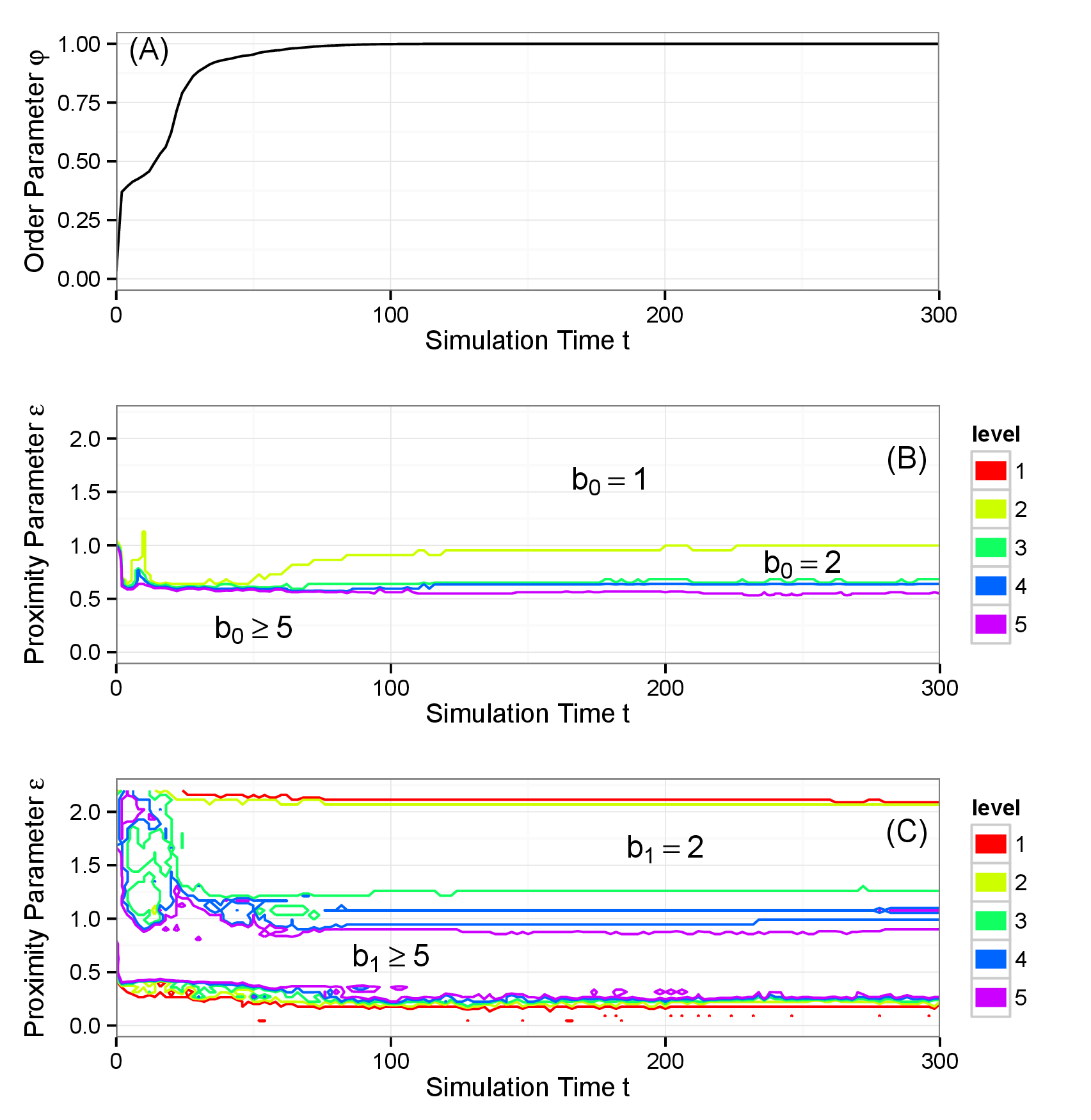}
\end{center}
\caption{\label{fig:vicsek1dTimeSeries}
{\bf Aggregate behavior of the Vicsek model, simulation \#3.} A typical snapshot is shown in Figure~\ref{fig:vicsekStates}(C). (A)  Normalized average velocity order parameter $\varphi(t)$. (B) Contour plot of Betti number $\betti_0(\proximity,t)$. (C) Contour plot of Betti number $\betti_1(\proximity,t)$. The topological analysis shows essentially no cluster formation; the narrow region in which $\betti_0(\proximity,t)=2$ arises from an isolated agent. The two persistent topological circles are consistent with highly aligned particles covering both dimensions of the periodic spatial domain. Topological features become fairly stagnant once the entire group forms a large, aligned cluster traveling as a rigid body early in the simulation. See text for a more comprehensive analysis.}
\end{figure}

Figure~\ref{fig:vicsek1dTimeSeries} shows results from the same simulation as the snapshot in Figure~\ref{fig:vicsekStates}(C). Here, $N=300$, $\ell=5$, and $\eta=0.1$. In panel (A), the order parameter $\varphi(t)$ rises steeply to one with a small intermediate step. The order parameter here and in Figure~\ref{fig:vicsek1bTimeSeries}(A) share some features, and yet we will see that the topological calculations capture the differences in the dynamics already apparent in the snapshots of Figure~\ref{fig:vicsekStates}. 

Figure~\ref{fig:vicsek1cTimeSeries}(B) shows $\betti_0(\proximity,t)$, measuring the number of connected components. In the large region below the purple (bottom) contour, $\betti_0(\proximity,t) \geq 5$, which we interpret as noise. In the large region at the top, above the yellow contour, $\betti_0(\proximity, t) = 1$, indicating one connected component at the largest scales. The most noticeable intermediate region is in between the yellow and green contours where $\betti_0(\proximity, t) = 2$, indicating a strong signal of two connected components at the scale $0.6 \leq \proximity \leq 1$. We return to a discussion of these two connected components after a discussion of $\betti_1(\proximity,t)$.

Figure~\ref{fig:vicsek1cTimeSeries}(C) shows $\betti_1(\proximity,t)$, measuring the number of topological circles. At $t \approx 20$, there is a marked transition to $\betti_1(\proximity,t) = 2$ in the upper portion of the contour diagram. Taking into account that the order parameter $\varphi(t)$ indicates alignment -- and thus the lack of a topological circle in the heading dimension -- the data is reduced to living on the two-torus of the periodic spatial domain. For the snapshot $t=300$, we have computed that $\betti_2(\proximity,300) = 0$ in the range of $\proximity$ where $\betti_1(\proximity,t) = 2$. This indicates there are no trapped volumes. The first three Betti numbers $\betti = (1,2,0)$ are those of a punctured torus, and this interpretation is consistent with the hole in the data seen in the upper right quadrant of Figure~\ref{fig:vicsekStates}(C).

We now return attention to the region of Figure~\ref{fig:vicsek1dTimeSeries}(B) in which $\betti_0(\proximity,t)=2$ over a range of $\proximity$, indicating two connected components. Once the group has achieved strong alignment, agents do not change their relative configuration, and the population travels as a rigid body. Thus, there is potential for any outliers to remain outliers terminally. This is indeed the case in our simulation. There is a single agent in Figure~\ref{fig:vicsekStates}(C) (second from the top along the right hand border) who is isolated. It is only at the scale of the distance to its nearest neighbor that the transition to $\betti_0(\proximity,t)=1$ occurs.

\subsection*{Summary of Vicsek Model Analysis}

For both the first and third simulations, the order parameter $\varphi(t)$ -- shown in Figures~\ref{fig:vicsek1bTimeSeries}(A) and ~\ref{fig:vicsek1dTimeSeries}(A) -- increases rapidly to $\varphi(t) \approx 1$ with little variation. However, the topological structures are distinct. The first simulation shows cluster formation and eventual coverage across one dimension of the spatial domain. The third simulation shows one cluster (with an outlier) and coverage across both dimensions of the spatial domain. Curiously, this is the same topological signature as the second simulation, even though the order parameter in \ref{fig:vicsek1cTimeSeries}(A) is noisy and achieves a lower degree of alignment. Thus, the persistent homology computations capture the fact that in some sense, the second and third simulations are topologically equivalent. Also, as previously discussed, these computations reveal dynamical events not discernible in the order parameter, and we find them to be a helpful complement.

\vspace{-0.0000001in}

\section*{Analysis of the D'Orsogna Model}
\label{sec:dorsogna}

Another model is that of D'Orsogna and collaborators \cite{DOrChuBer2006,ChuDOrMar2007}, cited over 300 times as of the writing of this manuscript, and itself a refinement of \cite{LevRapCoh2001}. In contrast to the alignment-driven Vicsek model, the D'Orsogna model hinges on attractive-repulsive interactions between particles and produces many patterns including rotating rings, traveling swarms, and vortex states -- sometimes called mills -- reminiscent of fish schools.

The D'Orsogna model is a continuous-time dynamical system that describes the motion of interacting point particles in an unbounded plane. The model takes the form of Newtonian force equations and thus is second order in time. The equations are
\begin{subequations}
\label{eq:dorsognaModel}
\begin{eqnarray}
\dot{\vec{x}}_i & = & \vec{v}_i,\\
m \dot{\vec{v}}_i & = & \left( \alpha - \beta |\vec{v}_i|^2 \right)\vec{v}_i - \nabla_i Q_i,\\
Q_i & = & \sum_{j \neq i} C_r \exp{-|\vec{x}_i-\vec{x}_j|/L_r} - C_a \exp{-|\vec{x}_i-\vec{x}_j|/L_a}.
\end{eqnarray}
\end{subequations}
Here, $\vec{x}_i(t) \in \mathbb{R}^2$ is the position of particle $i=1,
\ldots,N$ at time $t$, and $\vec{v}_i(t) \in \mathbb{R}^2$ is velocity. The first equation simply defines velocity as the derivative of position. The second equation is Newton's law, stating that mass times acceleration is equal to a sum of forces. These forces include self-propulsion of strength $\alpha$, friction of strength $\beta$, and interaction forces described by the potential $Q$. The first term in $Q$ describes repulsion of strength $C_r$ and characteristic length scale $L_r$. The second term is similar, but describes attraction of strength $C_a$ and characteristic length scale $L_a$. Put together, these two terms are similar to potentials used in molecular physics. In biological scenarios, typically $L_r < L_a$ and $C_r > C_a$, meaning that repulsion occurs over shorter distances and is stronger. For an isolated pair of particles interacting solely according to this attractive-repulsive rule, and for appropriately chosen parameters, the potential has a unique minimum, and there exists an equilibrium distance at which attraction and repulsion balance. When one deals with an ensemble of $N$ particles each experiencing pairwise interactions, the behavior is highly nontrivial. Restated in brief, (\ref{eq:dorsognaModel}) prescribes that each particle obeys Newton's law, with the relevant forces being self-propulsion, friction, and pairwise attraction-repulsion with all other particles.

Arguably, one of the most intriguing behaviors of the model is the formation of mills, occurring in certain parameter regimes. These structures are annular in shape, with particles rotating around a hollow core. In a single mill, all particles travel with the same orientation (clockwise or counterclockwise). In a double mill, some particles travel clockwise and some travel counterclockwise. It is helpful to think about the topology of these states. A mill and a double mill have distinct topologies in four dimensional position-velocity space. The single mill is one connected component and one topological circle, that is, $\betti = (1,1,0,\ldots)$. The double mill is two connected components each of which is a topological circle, that is, $\betti = (2,2,0,\ldots)$. The two circles are concentric, nonplanar, and nonintersecting in four dimensional space.

\begin{figure}[ph!]
\begin{center}
\includegraphics{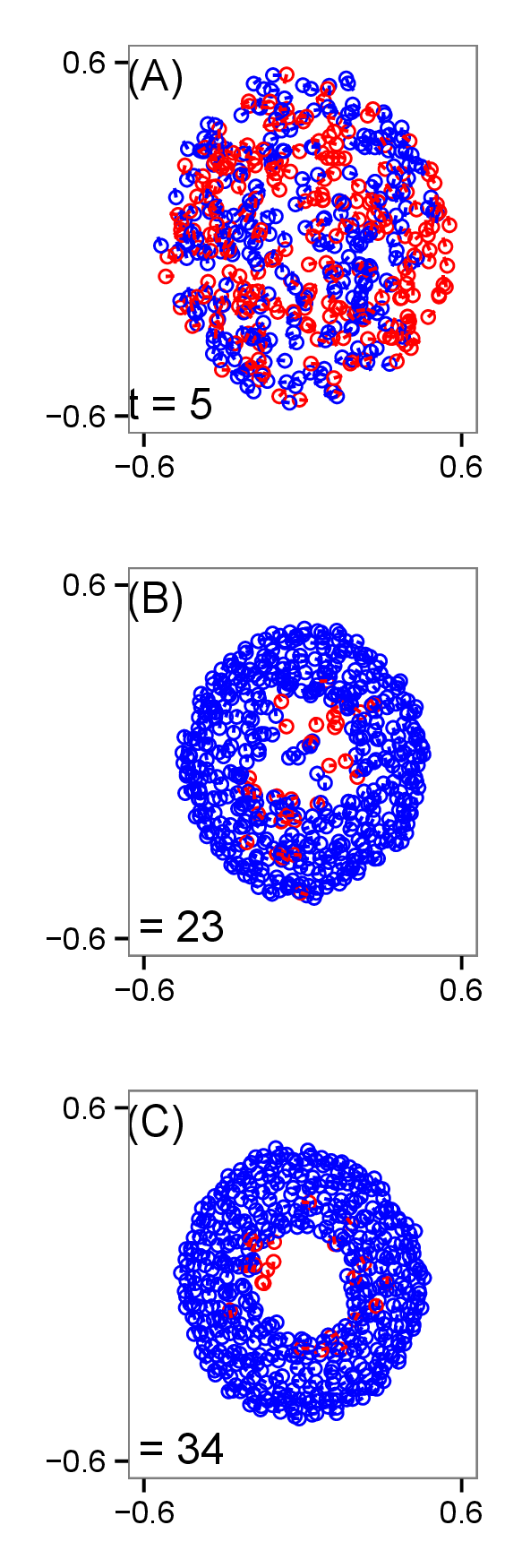}
\end{center}
\caption{\label{fig:dorsognaStates}
{\bf Simulation snapshots of the D'Orsogna model (\ref{eq:dorsognaModel}).} Circles indicate positions of the $N=500$ particles in an unbounded plane, line segments represent heading, and blue (red) agents are traveling (counter)clockwise. Over time, the group develops a hollow core and a double-mill structure in which a majority of agents travel clockwise, but a minority persists in the counterclockwise orientation. (A) Time $t=5$. (B) Time $t=23$. (C) Time $t=34$. The other model parameters used in this simulation are $\alpha = 1.5$, $\beta=0.5$, $C_r = 1$, $L_r = 0.5$, $C_a = 0.5$, $L_a = 2$.}
\end{figure}

We conduct a simulation of (\ref{eq:dorsognaModel}) with $N=500$ particles, and model parameters $\alpha = 1.5$, $\beta = 0.5$, $C_a = 0.5$, $C_r = 1$, $L_a = 2$, $L_r = 0.5$. Figure~\ref{fig:dorsognaStates} shows selected simulation snapshots. For convenience, we color blue all particles traveling clockwise with respect to the center of mass of the group; particles traveling counterclockwise are red. In panel (A), at $t=5$, the particles occupy a disk-like region with somewhat disorganized velocities. At $t=23$, we see that a hollow core has begun to form. At $t=45$, the mill structure has a well defined core. The group favors the clockwise direction, though there is a minority group of particles traveling counterclockwise.

\begin{figure}[ph!]
\begin{center}
\includegraphics[width=0.9\textwidth]{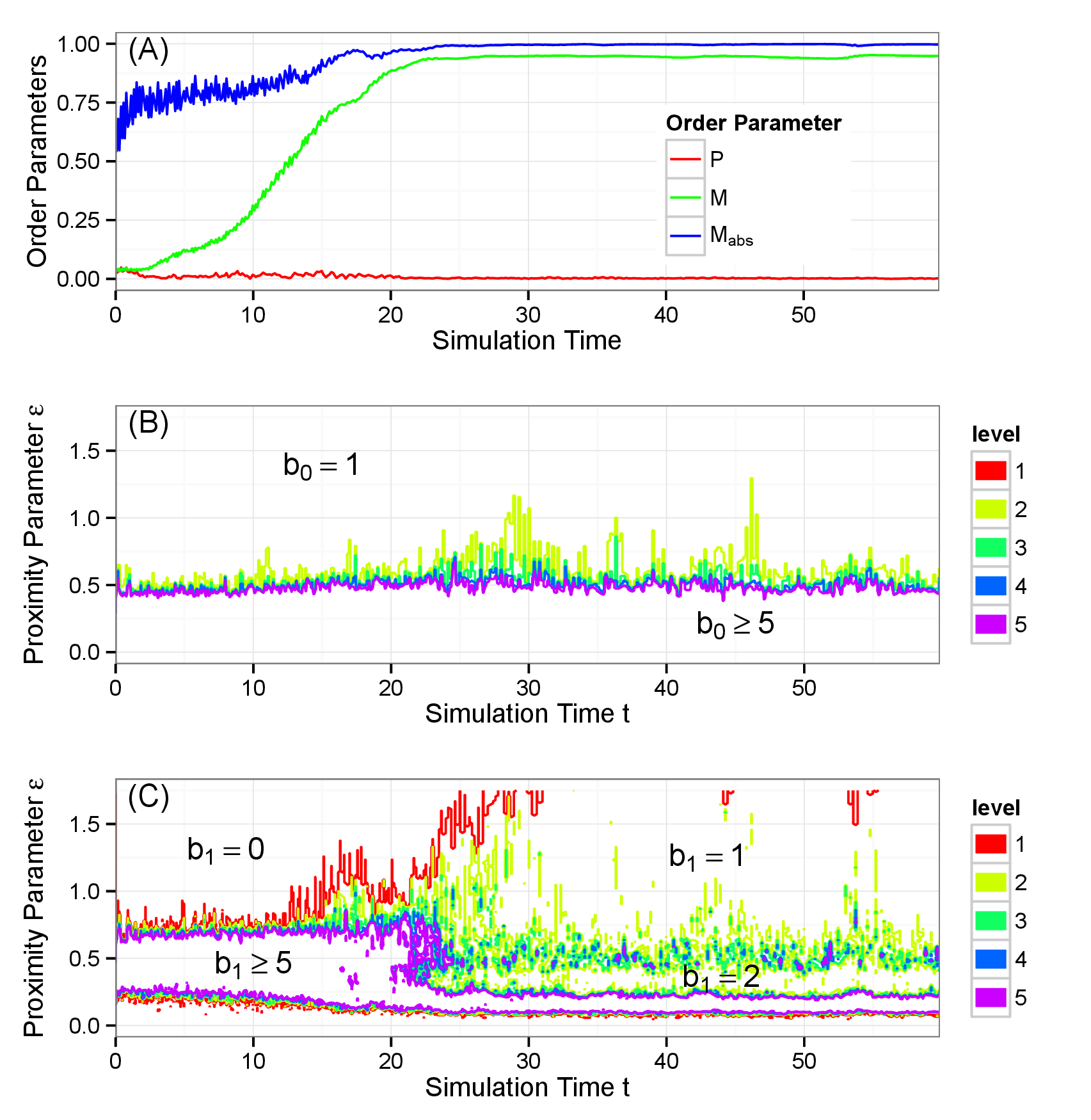}
\end{center}
\caption{\label{fig:dorsognaTimeSeries}
{\bf Aggregate behavior of the D'Orsogna model.} Snapshot of the time evolution are shown in Figure~\ref{fig:dorsognaStates}. (A) Three order parameters: polarization $P$ (red), angular momentum $M$ (green), and absolute angular momentum $M_{abs}$ (blue). (B) Contour plot of Betti number $\betti_0(\proximity,t)$. (C) Contour plot of Betti number $\betti_1(\proximity,t)$. At times below $t \approx 20$, there is little topological structure. For $t > 20$, we have one or -- intermittently -- two connected components of data points. There are two discernible topological circles for smaller $\proximity$ and one circle for larger $\proximity$. These circles survive for long periods of simulation time. The topological signature of the first two Betti numbers, $\betti = (2,2)$, is consistent with a double mill structure. See text for a more comprehensive analysis.}
\end{figure}

Figure~\ref{fig:dorsognaTimeSeries}(A) shows time series of three order parameter metrics used in~\cite{ChuDOrMar2007} to characterize the global behavior of the system. Polarization $P$, defined in (\ref{eq:metrics1}), is similar to $\varphi(t)$ for the Vicsek model; it measures the degree to which agents are aligned. Because the group is traveling around a circle, $P$ (red curve) remains low for the duration of the simulation. Angular momentum $M$, also defined in (\ref{eq:metrics1}), helps quantify the rotation of the group. For a perfect mill structure, $M=1$. For our simulation, we see an evolution from $M \approx 0$ early in time to $M \approx 0.93$ for later times (green curve). However, as mentioned previously, the metric $M$ cannot distinguish between single and double mills, and so \cite{ChuDOrMar2007} introduces the absolute angular momentum $M_{abs}$, defined in \eqref{eq:metrics2}. The fact that $M_{abs}$ approaches unity signals that the asymptotic behavior of the group is rotational. The fact that $M$ approaches a number close to but less than unity signals that a small minority of the group members are rotating counter to the majority.

Figure~\ref{fig:dorsognaTimeSeries}(B) shows $\betti_0(\proximity,t)$, measuring the number of connected components in the four dimensional space defined by position and velocity. In the large region below the purple (bottom) contour, $\betti_0(\proximity,t) \geq 5$. The connected components in this region persist only for very short ranges of $\proximity$; this is a noisy topological signal which we disregard. In the large region at the top, above the yellow contour, $\betti_0(\proximity, t) = 1$, indicating one connected component at the largest scales, that is, a component consisting of all the data points. The bottom and top regions in the graph are separated by a noisy boundary of intermediate curves. These curves indicate that at scales  approximately in the range $0.5 < \proximity < 1.0$ there is, over time, sporadic coagulation and fragmentation of connected components. For example, at times near $t \approx 30$ and $t \approx 46$, the region between the yellow and green contours is thicker, indicating a more persistent signal of two connected components.

Panel (C) shows $\betti_1(\proximity,t)$, measuring the number of topological circles in four dimensional space. In the bottom region of the graph enclosed by the purple contour, $\betti_1(\proximity,t) \geq 5$ which, as before, we interpret as noise. At early times, the only non-noisy signal is the large $\betti_1(\proximity,t)=0$ region above the red contour, indicating an absence of topological circles. Starting at $t \approx 20$, a marked transition occurs, and becomes persistent by $t \approx 30$. This transition detects the formation of a hole in the simulation, as shown in Figure~\ref{fig:dorsognaStates}(B). For $t \geq 30$, there is a discernible region of the contour diagram in which $\betti_1(\proximity,t) = 2$, indicating two topological circles in four dimensional position-velocity space. For these same times, over larger values of $\proximity$, $\betti_1(\proximity,t) = 1$, indicating the loss of a circle. A topological circle could disappear by closing across its diameter or by merging with another circle. For our data, antipodal points in the mill with opposite orientations of travel are approximately $0.5$ units apart (as seen in Figure~\ref{fig:dorsognaStates}) and this is (approximately) the proximity scale at which $\betti_1$ transitions from two to one, indicating that the two mills merge into one at this scale. The remaining topological circle will be lost when it closes on itself across its diameter for sufficiently large values of $\proximity$ (not shown).

Pulling together the information from panels (B) and (C), we conclude the following. At times below $t \approx 20$, there is little topological structure. Then, a clear topological transition occurs. For later times, we have one or -- intermittently -- two connected components of data points. There are two discernible topological circles for smaller $\proximity$ and one circle for larger $\proximity$. These circles survive for long periods of simulation time. The topological signature of the first two Betti numbers, $\betti = (2,2)$, is consistent with a double mill structure. The noisiness of the second connected component arises from the sparsity of the counterclockwise mill in four dimensional space.

\section*{Conclusions}
\label{sec:conclusions}

Inspired by physics, order parameters such as polarization and angular momentum have been useful for characterizing the global behavior of biological aggregations. We propose topological data analysis as an additional, valuable technology for understanding their group behavior.

We have performed numerical simulations of two well-known mathematical models of biological aggregations, resulting in point clouds of data that evolve in time. To understand the global behavior of each model, we study the topological structure of the point clouds by calculating their persistent homology. More specifically, we compute Betti numbers, which count connected components, topological circles, trapped volumes, and so forth.

To interpret the topological computations, we introduce a new visualization tool, namely a Contour Realization Of Computed $k$-dimensional hole Evolution in the Rips complex (CROCKER), which track Betti numbers across both proximity scale and simulation time. In topological data analysis, persistent features in a static point cloud correspond to long bars in a topological barcode. In our analysis, features persisting over scale and simulation time appear as large regions in the contour plot.

In Vicsek's model of aligning particles, the homological measures distinguish simulations that the usual alignment order parameter cannot. They also find topological similarity between simulations with different order parameter time series. In D'Orsogna's model of self-propelled, attracting-repelling particles, the topological calculations recognize the presence of a double mill state. In our study we have, for tutorial purposes, sought to explain our CROCKER plots by a subsequent manual examination of the data. That said, though phenomena such as group alignment, clustering, and double mills could be seen upon detailed examination of our raw simulation data, we would not have found them by eye if the topological methods had not first detected them.

One limitation of our work is that we have only calculated the first two Betti numbers, $\betti_0$ and $\betti_1$, except for a small number of isolated cases in which we have also calculated $\betti_2$. Calculating higher Betti numbers of our point clouds would yield additional information, but is computationally costly. Another limitation is that topological persistence over scale is different from persistence over time. For a fixed simulation time, the topological barcode is guaranteed to measure the \emph{same} topological features through multiple proximity scales because nested sequences of simplicial complexes form a filtration over which homology persists. However, this guarantee does not hold over simulation time. For example, if $\betti_0 = 4$ indicating four connected components in two successive frames of a simulation, there is no mathematical guarantee that these are the same four connected components. Nonetheless, because the aggregation models we study evolve smoothly in time, we expect persistent topological features to do so as well.

One attempt to address time evolution of topological features uses vineyards, which have been applied to protein folding in the context of level set persistence \cite{CohEdeMor2006}. Another attempt might involve multidimensional persistence, which allows a persistence computation simultaneously over multiple parameters \cite{CarZom2009}. It could be useful to apply these two tools to biological aggregations. It could also be useful to consider another topological approach, namely braids, which have yielded insight into other dynamical systems applications such as fluids and crowd dynamics \cite{AllThi2012,Ali2013}. Finally, our main goal has been to demonstrate the utility of topological data analysis for biological aggregations and similar applications. We have used the Vicsek and D'Orsogna models as convenient examples, and focused on a small number of simulations. That said, it could be revealing to conduct large numbers of randomly-seeded simulations for fixed parameters, compute the persistent homology of each one, and average this topological data. Doing so would allow more precise quantification of the timescales and persistence scales of the topological transitions.

Topological data analysis is an active and growing area of current research. We hope that our work above contributes to the toolkit that applied mathematicians might bring to bear on models they study.

\bibliography{master_bibliography}

\end{document}